\documentstyle[aps,prb,floats,epsfig,twocolumn]{revtex}

\begin{document}
\draft
\wideabs{
\title{SDW and FISDW transition of (TMTSF)$_2$ClO$_4$ at high magnetic fields}
\author{N. Matsunaga}
\address{CRTBT-CNRS, laboratoire associ\'{e} \`{a} l'UJF, BP 166, 38042 Grenoble Cedex 9, France\\
and Division of Physics, Hokkaido University, Sapporo 060-0810, Japan}
\author{A. Briggs}
\address{CRTBT-CNRS, laboratoire associ\'{e} \`{a} l'UJF, BP 166, 38042 Grenoble Cedex 9, France}
\author{A. Ishikawa and K. Nomura}
\address{Division of Physics, Hokkaido University, Sapporo 060-0810, Japan}
\author{T.Hanajiri, J. Yamada, S. Nakatsuji and H. Anzai}
\address{Department of Material Science, Himeji Institute of Technology, Kamigohri 678-1297, Japan}
\date{Submitted March 30, 2000}
\maketitle
\begin{abstract}
The magnetic field dependence of the SDW transition 
in (TMTSF)$_2$ClO$_4$ for various anion cooling rates has been 
measured, with the field up to 27T parallel to the lowest 
conductivity direction $c^{\ast}$.
For quenched (TMTSF)$_2$ClO$_4$, the SDW transition temperature $T_{\rm {SDW}}$ increases from 4.5K 
in zero field up to 8.4K at 27T. 
A quadratic behavior is observed below 18T, 
followed by a saturation behavior. These results are consistent 
with the prediction of the mean-field theory.
From these behaviors, $T_{\rm {SDW}}$ is estimated as $T_{\rm {SDW_0}}$=13.5K 
for the perfect nesting case.
This indicates that the SDW phase in quenched (TMTSF)$_2$ClO$_4$,
 where $T_{\rm {SDW}}$ is less than 6K,
is strongly suppressed by the two-dimensionality of the system.
In the intermediate cooled state in which the SDW phase does not appear
in zero field,  
the transition temperature for the field-induced SDW shows a quadratic behavior above 12T 
and there is no saturation behavior even at 27T, in contrast to the 
FISDW phase in the relaxed state.
This behavior can probably be attributed to the difference of the dimerized gap due to anion ordering.

\end{abstract}
%\pacs{PACS numbers: 75.30.Fv, 72.15.Gd, 74.70.Kn}
%Do not delete this line
} 
\narrowtext

The compounds (TMTSF)$_2$X, where TMTSF denotes tetramethyltetraselenafulvane 
and X=PF$_6$, AsF$_6$, ClO$_4$ etc., are highly anisotropic organic conductors 
with a rich phase diagram exhibiting spin-density-wave (SDW), metallic, 
superconducting and field-induced SDW (FISDW) phases at low temperatures \cite{Ishiguro}.
At ambient pressure the PF$_6$ salt undergoes a metal-SDW transition 
at 12K.
The temperature of the SDW transition decreases with pressure and ultimately tends to zero.
Above a critical pressure, the superconducting phase appears at $T_{\rm C}$=1.1K.
In the magnetic field parallel to the lowest conductivity direction $c^{\ast}$ and above the critical pressure, 
a cascade of FISDW states is observed, with quantized Hall resistance $\rho_{xy} \sim h/(n2e^2)$ in the 
sequence n=.....4,3,2,1,0 as the magnetic field is increased.
The states labeled with integer n have been identified as semimetallic FISDW states
while that with n=0 is a FISDW insulating state.
The phase diagram of the metallic-SDW and FISDW transition 
for the PF$_6$ salt is successfully explained 
by the mean field theory based on the nesting of the slightly warped,
quasi one-dimensional Fermi surface. 

In (TMTSF)$_2$ClO$_4$, the ordering of the non-centrosymmetric anion ClO$_4^-$
occurs at an anion ordering temperature $T_{\rm {AO}}\sim $24K.
For the ClO$_4$ salt slowly cooled (relaxed) through $T_{\rm {AO}}$, the anion ordering 
creates a superlattice potential with a wave vector Q=(0,1/2,0) \cite{Pouget}
and the superconducting phase appears below 1.2K.
When the sample is rapidly cooled (quenched) through $T_{\rm {AO}}$, 
the orientations of anions are frozen in random directions and the SDW phase is induced below 6K.
With decreasing cooling rate, the SDW transition 
temperature $T_{\rm {SDW}}$ decreases \cite{Schwenk}. 
This phenomenon is similar to the results for (TMTSF)$_2$PF$_6$, 
if we identify the effect of pressure with that of cooling rate. 

Within the standard model based on the imperfect nesting theory, in the
SDW phase the closed orbits region due to the two-dimensionality of the
system is near the SDW gap created in the band structure.
When the magnetic field is applied along the c-axis,
the total energy of the SDW proves to be lowered by quantization of these closed orbits,
since the zero-field state density in the energy region of closed-orbits is a 
decreasing function of energy \cite{Ishiguro}.
As a result of this energy gain, it has been predicted that $T_{\rm {SDW}}$
increases nearly quadratically with the field in low magnetic field,
as $T_{\rm {SDW}}(H)=T_{\rm {SDW}}(0)+{\rm C}H^2$ where C is a constant,
and shows a saturation behavior in high magnetic field. 
The quadratic magnetic field dependence has been confirmed by the experiments for 
(TMTSF)$_2$PF$_6$ \cite{Danner} and quenched (TMTSF)$_2$ClO$_4$ \cite{Mat-ClO4}.

While relaxed (TMTSF)$_2$ClO$_4$ exhibits superconductivity and 
a cascade of FISDW, similar to the behavior observed 
in (TMTSF)$_2$PF$_6$ above the critical pressure, some experimental results 
for relaxed (TMTSF)$_2$ClO$_4$ cannot be explained in terms of 
the standard model which works so well for the PF$_6$ salt i.e.
(1)In the low-field cascade of FISDW's the sequence of Hall plateaus
is not in the expected order \cite{Ribault}.
(2)The n=0 insulating state is not observed, but a very stable quantum Hall
state is observed from 7.5 to 27T \cite{Chambemberlin,Noughton}.
(3)The FISDW transition temperature
is about 5.5K and is essentially independent of field above 15T \cite{McKernan}.
For the ClO$_4$ salt slowly cooled through $T_{\rm {AO}}$, the anion ordering 
creates a superlattice potential and it separates 
the original Fermi surface into two pairs of open sheets.
This dimerized gap due to anion ordering is usually thought to be the cause of above anomaly. 

In this paper, we describe the cooling rate dependence 
of the SDW and FISDW transition for (TMTSF)$_2$ClO$_4$ under the strong magnetic field.
We aimed:
(1) to check the mean-field predictions 
and to estimate an absolute value of the characteristic energy
for the SDW state of quenched (TMTSF)$_2$ClO$_4$;
(2) to confirm a new phase diagram which 
has been proposed for the FISDW relaxed state;
(3) to check the role of the dimerized gap due to anion ordering
in the FISDW state.
\begin{figure}
\epsfxsize=2.7in \center \epsfbox{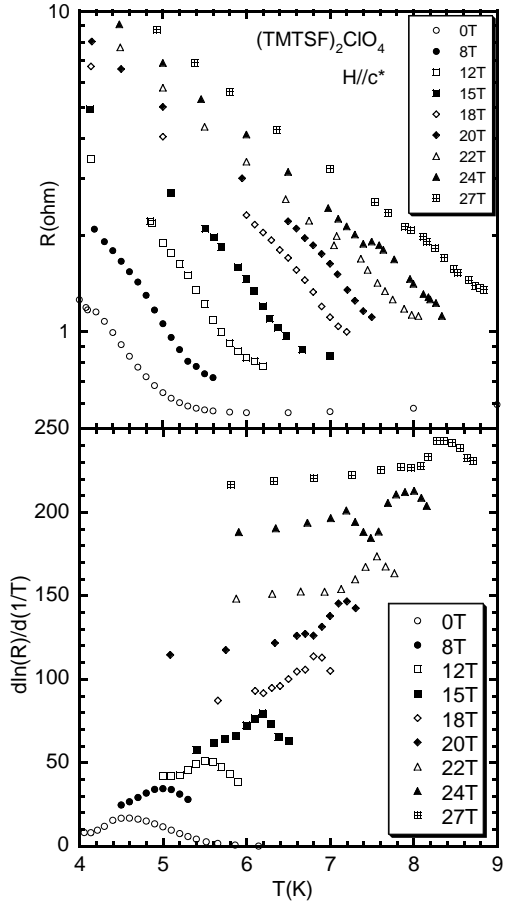} \vspace{0.1in}
\caption{(Top) Temperature dependence of the transverse magnetoresistance in quenched 
(TMTSF)$_2$ClO$_4$ at constant field along $c^{\ast}$ axis as indicated. 
(Bottom) Logarithmic derivatives of the above data
offset from zero for clarity.} \label{fig:1}
\end{figure}

Single crystals of (TMTSF)$_2$ClO$_4$ were synthesized by the standard electro-chemical method.
The resistance measurements along the conducting a-axis
 were carried out using a standard four probe AC method  
over the temperature range from 1.2K to 10K. 
Electric leads of 10$\mu$m gold wire were attached with silver paint onto 
gold evaporated contacts.
The quenching of the sample was done by immersing it into liquid helium directly 
after annealing at 30K.
The temperature was measured using a Cernox CX-1050-SD
resistance thermometer.
The measurements in the fields to 27T were done 
in a resistive magnet at the Grenoble High Magnetic Field Laboratory.

Figure 1 shows the temperature dependence of the transverse magnetoresistance
with the field parallel to the lowest conductivity direction $c^{\ast}$ in quenched 
(TMTSF)$_2$ClO$_4$, in which the cooling rate is faster than 10K/s.
From this figure it is clear that the transition is shifted 
toward higher temperature as the field is increased.
The derivative of the logarithm of these data with 1/T is also shown in Fig.1.
The SDW transition temperatures $T_{\rm {SDW}}$ determined from the peak
in the derivative are plotted vs field in Fig.2.
For quenched (TMTSF)$_2$ClO$_4$, 
$T_{\rm {SDW}}$ increases from 4.5K in zero field up to 8.4K at 27T.
As predicted in the mean-field theory \cite{Montambaux},
 a quadratic behavior is observed below 18T, followed by a saturation behavior.
The solid line is the best fit of the quadratic function
 to experimental data in the low field range
and is described by $T_{\rm {SDW}}(H)=4.58+0.0070H^2$.
This coefficient of the quadratic term is consistent with previous results \cite{Mat-ClO4}.
\begin{figure}
\epsfxsize=2.6in \center \epsfbox{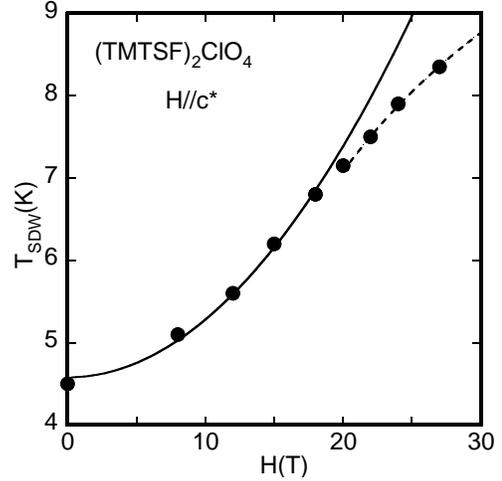} \vspace{0.1in}
\caption{Magnetic field dependence of the SDW transition temperature $T_{\rm {SDW}}$ for quenched 
(TMTSF)$_2$ClO$_4$.
The full line is the best fit of the quadratic function.
The dashed line is the fit of mean field theory.} \label{fig:2}
\end{figure}
From the observed curvature of the field dependence of $T_{\rm {SDW}}$ in high magnetic field, 
we can determine several important parameters 
for the SDW transition using a mean-field theory \cite{Maki,Bjelis}.
The relevant parameters entering this theory are:
the SDW transition temperature corresponding 
to the perfect nesting case $T_{\rm {SDW_0}}$, Fermi velocity $v_{\rm F}$ 
and the ratio of the energy parameter $\epsilon_0$ 
which characterizes the deviation from perfect nesting
to the SDW order parameter $\Delta_0$ at T=0K for $\epsilon_0$=0.
When  $\epsilon_0$ becomes equal to  $\Delta_0$ (i.e., $\epsilon_0$/$\Delta_0$=1), 
the nesting of the Fermi surface is so imperfect that the SDW state is suppressed.
The dashed line in Fig.2. is the fit of this theory to the experimental data.
The agreement is good.
The fit gives $v_{\rm F}$=7.8$\times$10$^4$m/s, $T_{\rm {SDW_0}}$=13.5K
and $\epsilon_0$/$\Delta_0$=0.994 using a lattice parameter along the $b^{\prime}$ axis of 7.7$\times$10$^{-10}$m.
This indicates that the mean field calculations nicely account for the field induced increase of 
$T_{\rm {SDW}}$ and the SDW phase of quenched (TMTSF)$_2$ClO$_4$, where $T_{\rm {SDW}}$
is less than 6K, is strongly suppressed 
by the two-dimensionality of the system.

Measurements of the temperature dependence of the transverse magnetoresistance 
in relaxed (TMTSF)$_2$ClO$_4$, in which the cooling rate is about 0.001K/s, 
are shown in Fig.3.
The FISDW transition is observed above 12T.
In the case of relaxed (TMTSF)$_2$ClO$_4$, the increase of resistance starts sharply 
at the transition temperature in contrast to the case of the quenched and intermediate cooled states.
The FISDW transition temperatures $T_{\rm {FISDW}}$ of relaxed (TMTSF)$_2$ClO$_4$ 
determined from the intersection of 
the extrapolations of the resistivity curve from the FISDW 
and metallic phases are plotted vs field in Fig.4 (open squares).
For relaxed (TMTSF)$_2$ClO$_4$, 
$T_{\rm {FISDW}}$ increases from 4.5K at 12T to 5.5K at 18T and is almost independent of field above 18T.
This result supports the new phase diagram which 
has been proposed for the relaxed (TMTSF)$_2$ClO$_4$ \cite{McKernan}.
\begin{figure}
\epsfxsize=2.8in \center \epsfbox{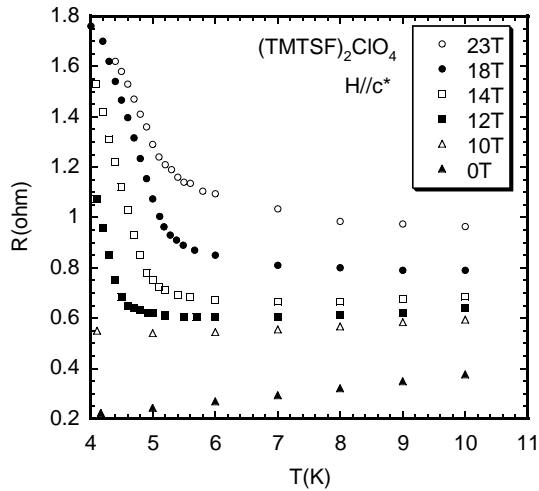} \vspace{0.1in}
\caption{Temperature dependence of the transverse magnetoresistance in relaxed 
(TMTSF)$_2$ClO$_4$ at constant field along $c^{\ast}$ axis as indicated. } \label{fig:3}
\end{figure}
\begin{figure}
\epsfxsize=2.6in \center \epsfbox{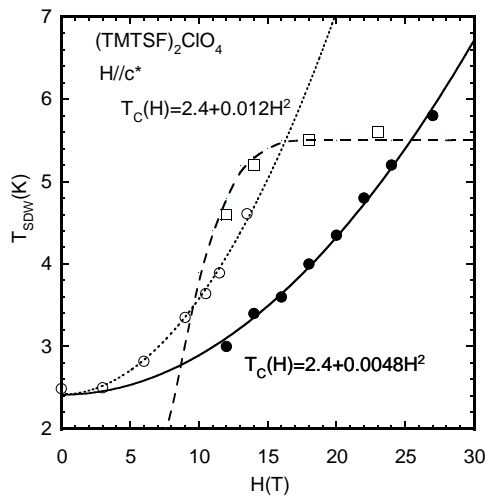} \vspace{0.1in}
\caption{Magnetic field dependence of the SDW transition temperature $T_{\rm {SDW}}$
for the relaxed (open squares), intermediate cooled (full circles) and rapidly cooled (open circles) state.
The dashed line is the phase boundary for the relaxed state proposed by McKernan {\it et al}  \cite{McKernan}.
The full and dotted lines are the best fit of the quadratic function 
for the intermediate and rapidly cooled states, respectively. } \label{fig:4}
\end{figure}

The temperature dependence of the transverse magnetoresistance
in the intermediate cooled (TMTSF)$_2$ClO$_4$, in which the cooling rate is about 0.5K/s, is shown in Fig.5.
In this state, there is also no SDW phase in zero field \cite{mat1} 
and the FISDW state appears above 12T at 3K \cite{mat2}.
It is clear that the transition temperature is shifted toward higher temperatures as the field is increased.
The derivative of the logarithm of these data with 1/T is also shown in Fig.5.
The transition temperatures $T_{\rm {FISDW}}$ determined from the peak in the derivative
are plotted vs field in Fig.4 (full circles).
As shown in Fig.4, $T_{\rm {FISDW}}$ shows a quadratic behavior above 12T 
and there is no saturation even at 27T.

\begin{figure}
\epsfxsize=2.7in \center \epsfbox{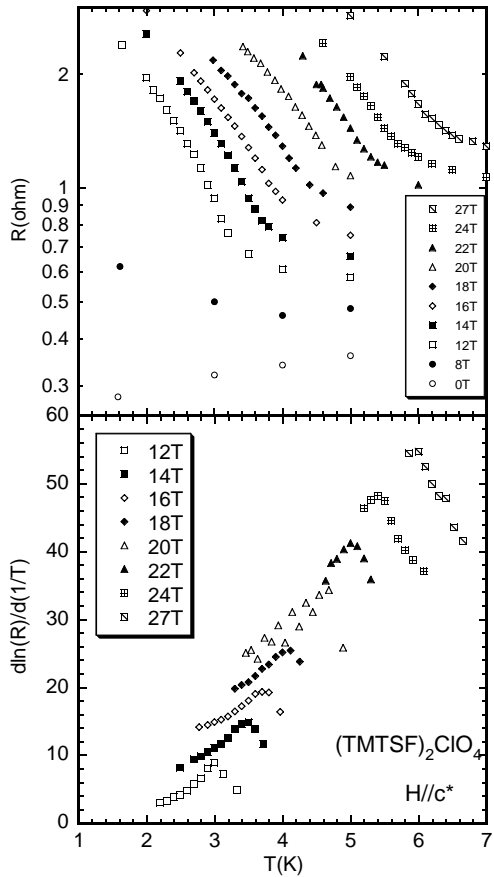} \vspace{0.1in}
\caption{(Top) Temperature dependence of the transverse magnetoresistance in the intermediate cooled state of
(TMTSF)$_2$ClO$_4$ at constant field along $c^{\ast}$ axis as indicated. 
(Bottom) Logarithmic derivatives of the above data
offset from zero for clarity.} \label{fig:5}
\end{figure}

When the SDW transition appears in zero field,
$T_{\rm {SDW}}$ decreases and the coefficient of the quadratic term increases 
with decreasing cooling rate \cite{Mat-ClO4}.
These results are consistent with the prediction of the mean field theory \cite{Montambaux}.
In the intermediate cooled state,
$T_{\rm {FISDW}}$ also shows a quadratic behavior above 12T,
but there is no SDW phase in zero field in contrast to the quenched state.
The value of $T_{\rm {FISDW}}$ at zero field extrapolated 
from the quadratic behavior is 2.4K as shown in Fig.4.
In our experiment on (TMTSF)$_2$ClO$_4$, 
we find the SDW state in which the SDW transition appears at 2.4K in zero field 
for a more rapidly cooled state with the cooling rate about 2K/s \cite{Mat-ClO4}.
The field dependence of $T_{\rm {SDW}}$ is shown in Fig.4(Open circles) and 
is also described by the quadratic behavior.
Although $T_{\rm {SDW}}$ for this SDW state at zero field is almost the same as
the extrapolated values of $T_{\rm {FISDW}}$ for the intermediate cooled state,
the coefficient of the quadratic term in the former SDW state 
is about twice as large as that in the latter.
%This result is not consistent with the prediction of the mean field theory that
%$T_{\rm {SDW}}$ decreases and the coefficient of the quadratic term increases 
%with increasing two-dimensionality of the system due to the decrease of cooling %rate.
The saturation of the quadratic term for the intermediate cooled state 
will occur for higher fields, because the coefficient of the quadratic term
becomes small. 
 
The X-ray diffraction studies indicate that the size of the regions, over which the anion is ordered, 
decreases with increasing cooling rate \cite{Kagoshima}.
Thus, the concentration of the scattering centers associated with 
the boundaries between anion ordered states increases with increasing cooling rate.
In fact, the residual resistances in the quenched and the intermediate cooled state are about 2 and 1.5 times
that in the relaxed state around 6K, respectively.
Because the periodic anion potential is out of phase at the boundary between anion ordered regions, 
the boundary not only works as a scattering center but also suppresses the anion gap 
in the intermediate cooled state, where the size of the ordered region is quite small.
As a result, the difference 
between the relaxed and intermediate cooled states can be explained by the assumption that the FISDW phase 
of the relaxed state is a peculiar state in (TMTSF)$_2$ClO$_4$ stabilized by the dimerized gap 
due to anion ordering and the FISDW phase of the intermediate cooled state is the standard FISDW state as observed 
in (TMTSF)$_2$PF$_6$ turned back on by the disappearance or a strong decrease of the anion gap. 
In fact, the T-H phase diagram of the intermediate cooled (TMTSF)$_2$ClO$_4$ looks very similar that of 
the FISDW phase in (TMTSF)$_2$PF$_6$ \cite{Hannahs}.
It seems reasonable that the quadratic increase of $T_{\rm {FISDW}}$ with magnetic field 
in the intermediate cooled state is not the quadratic increase predicted from the mean field theory 
in the SDW phase but the envelope line of the FISDW states with different quantum number n.

In summary, we have measured the magnetic field dependence of the SDW transition 
in (TMTSF)$_2$ClO$_4$ for various anion cooling rates with the field up to 27T and
shown that the mean-field calculations nicely account for the field-induced increase
of $T_{\rm {SDW}}$ in quenched (TMTSF)$_2$ClO$_4$.
Satisfactory values of both $T_{\rm {SDW_0}}$ and Fermi velocity have been obtained from the analysis.
In the intermediate cooled state in which the SDW phase does not appear
in zero field, $T_{\rm {SDW}}$ shows a quadratic behavior above 12T 
and there is no saturation even at 27T.
This behavior is quite different to the 
FISDW phase of the relaxed state.
The difference between the relaxed and intermediate cooled states can be explained by the assumption 
that the FISDW phase of the relaxed state is a peculiar state
stabilized by the dimerized gap due to anion ordering and
the FISDW phase of the intermediate state is 
the standard FISDW state
turned back on by the disappearance or a strong decrease of the anion gap.

Some of this work was carried out as part of the
'Research for the Future' project, JSPS-RFTF97P00105,
supported by the Japan Society for the Promotion 
of Science.

Note added-After submission of this article we were informed of work
a reporting similar magnetic field dependence of $T_{\rm{SDW}}$ for
quenched (TMTSF)$_2$ClO$_4$ \cite{Mielke}.

\end{document}